\documentclass[12pt]{article} \topmargin=0.5 cm  \textwidth 16.5
cm \textheight 23 cm \oddsidemargin 0pt \evensidemargin 0pt
\usepackage {graphicx}
\begin{document}

\title{Extra-Dimensions effects on the fermion-induced quantum energy in the presence of a constant magnetic field}
\author{K. Farakos \footnote{kfarakos@central.ntua.gr} and P. Pasipoularides \footnote{paul@central.ntua.gr}
\\ \\  Department of Physics, National Technical University of
       Athens \\ Zografou Campus, 157 80 Athens, Greece}
\date{ }
       \maketitle
\begin{abstract}
We consider a $U(1)$ gauge field theory with fermion fields (or with
scalar fields) that live in a space with $\delta$ extra compact
dimensions, and we compute the fermion-induced quantum energy in the
presence of a constant magnetic field, which is directed towards the
$x_{3}$ axis. Our motivation is to study the effect of extra
dimensions on the asymptotic behavior of the quantum energy in the
strong field limit ($eB>>M^{2}$), where $M=1/R$. We see that the
weak logarithmic growth of the quantum energy for four dimensions,
is modified by a rapid power growth in the case of the extra
dimensions.
\end{abstract}

\section{Introduction}

The computation of the fermion-induced quantum energy in the
presence of a constant magnetic field (or Heisenberg-Euler
lagrangian), is a topic that has attracted the attention of authors
from the early time of quantum electrodynamics \cite{1,2,3}. In
addition, the case of three dimensions has been studied in Ref.
\cite{4}. It is worth mentioning that inhomogeneous magnetic fields
have also been studied, analytically and numerically \cite{6,7,8,9}.
Finally we note that approximation tools such as the derivative
expansion are also available \cite{11}.

As it is believed, particle field theories like Standard Model are
embedded in more fundamental field theories, which may be string
theories. It is well known that string theories are formulated in
higher dimensional manifolds. For this reason, in recent years,
there has been a great interest for particle models with extra
compact dimensions (see for example \cite{A11,D11,12,B12,Z12,13}).

In the framework of the above discussion it would be interesting to
reconsider classical topics, like Heisenberg-Euler Lagrangian (for
QED), in the case of models with extra dimensions. The simplest way
to extend QED (or a $U(1)$ gauge field theory with fermions) in this
direction is to add one or more extra compact dimensions with radius
$R=1/M$, assuming that both the gauge and the fermionic fields live
in the bulk. We note that $M$ is the mass scale of the Kaluza-Klein
modes.

In this work we will generalize the well known Schwinger formula
\cite{3} for the effective energy in the case of $\delta$ extra
compact dimensions. Our motivation is to study the effect of the
extra dimensions on the strong field ($eB>>M^{2}$) asymptotic
behavior of the quantum part \footnote{The effective energy
$E_{eff}$ is equal to the classical energy $E_{class}$ plus a
quantum energy part $E_{Q}$, which is induced by the fermions, or
$E_{eff}=E_{class}+E_{Q}$.} of the effective energy. We see that the
weak logarithmic growth of the quantum energy, in the case of four
dimensions, is modified by a rapid power growth (see Eqs. (37),(43)
and (45) below) in the case of the extra dimensions. The reason is
that in the strong field limit ($eB>>M^{2}$) the extra dimensions
lose their compact structure and behave as if they were noncompact.

Previous works aimed at the study of extra noncompact dimensions
with external magnetic fields can be found in Ref. \cite{extra1}.
However, the topics which are covered in these works are different
to what this paper aims at.

The question that arises is weather the above mentioned model can be
assumed as an extra-dimension extension of QED. Note that the
smallest scale of extra dimensions that has been assumed is $M\sim
1Tev$ (for a specification of bounds on $M$ see Ref. \cite{14}),
according to the scenario of Refs. \cite{A11,D11,12}. It is obvious
that for $\sqrt{eB}\sim \; 1Tev$ (or $B\sim 10^{26}G$) the QED is
not valid. Thus, we will use this model for an understanding of the
effects of extra dimensions to the effective energy, but we can not
use it in order to extract trustworthy physical results (see also
the discussion in conclusions).

\section{Quantum energy for five dimensions}

In this section we study the case of five dimensions, and later we
will generalize our results for more extra dimensions.

The partition function of a U(1) gauge field theory with fermions,
in dimension D=5, reads
\begin{equation}
Z=\int_{b.c} {\cal D} A {\cal D}\bar{\Psi} {\cal D} \Psi e^{i \int
d^{D}x(-\frac{1}{4}F_{\mu\nu}F^{\mu\nu}+\bar{\Psi}(i \: \not \!
D-m_{f})\Psi+{\cal L}_{hd})}
\end{equation}
where $\int d^{D}x=\int_{0}^{2 \pi R}dx_{4}\int d^{4}x \:$ and
$x_{4}$ is the extra dimension, which is assumed to be compactified
on a circle of radius $R=1/M$ ($M$ is the scale of the Kaluza-Klein
modes).

In addition, we have assumed periodic boundary conditions for the
fermion and gauge fields, namely $\Psi(x,x_{4})=\Psi(x,x_{4}+2 \pi
R)$ and $A_{\mu}(x,x_{4})=A_{\mu}(x,x_{4}+2 \pi R)$, where
$x=(x_{0},x_{1},x_{2},x_{3})$.

At this point we remind the reader that the model we examine is
nonrenormalizable. However, in this paper, we will assume the above
model as a low energy effective field theory which is valid up to
large physical cut-off $\Lambda_{Ph}$, above which a new well
defined theory emerges. In particular, we assume that this effective
field theory has been obtained from an original fundamental field
theory by integrating out all higher momenta and heavy particles
above the physical cut-off $\Lambda_{Ph}$. Thus, the model we
examine is viewed as a Wilsonian effective field theory
\cite{peskin:14,Houang:13,Wilson:1}. For more details and criticism
on effective field theories with extra compact dimensions, see Ref.
\cite{Wilson:1}.

The lagrangian ${\cal L}_{hd}$ incorporates all the possible higher
dimension operators with dimension $2+D$ or higher, which respect
the symmetries of the above model. For example the lagrangian ${\cal
L}_{hd}$ should be Gauge invariant. However, the precise form of
${\cal L}_{hd}$ can not be determined, as the original
renormalizable fundamental theory that our model has come from, is
not known. In general, we will ignore the contributions of the
higher dimension operators, or we will drop the lagrangian ${\cal
L}_{hd}$ from the path integral of Eq. (1). However, in section 4.2,
it is necessary to accept the existence of a $2D$ dimension operator
of the form $w_{D}/2\int d^{D}x (F_{\mu\nu}F^{\mu\nu})^{2}$, in
order to incorporate the cut-off dependent part of the quantum
energy for $D=8,9,10$ (for details see section 4.2).

The effective action $S_{eff}[A]$ is defined by the equation
\begin{equation}
e^{i S_{eff}[A]}=\int_{b.c} {\cal D}\bar{\Psi} {\cal D} \Psi
e^{i\int d^{D}x \: (-\frac{1}{4}F_{\mu\nu}F^{\mu\nu}+\bar{\Psi}(i \:
\not \! D-m_{f})\Psi)}=e^{i\left(-\frac{1}{4}\int d^{D}x \:
F_{\mu\nu}F^{\mu\nu}+S_{Q}[A]\right)}
\end{equation}

The quantum part of the effective action $S_{Q}[A]$ is obtained by
integrating out the fermionic degrees of freedom in the path
integral, so we obtain
\begin{equation}
S_{Q}[A]=\frac{1}{i}\ln\int {\cal D}\bar{\Psi} {\cal D}\Psi e^{i\int
d^{D}x\: \bar{\Psi}(i \: \not \! D-m_{f})\Psi}=\frac{1}{i}Tr\ln(i
\not\!\!D-m_{f})
\end{equation}
where  $\not\!\! D=\Gamma^{\mu}(\partial_{\mu}-ie_{5}A_{\mu})$
$(\mu=0,1,2,3,4)$, and $e_{5}$ is the five dimensional coupling
constant. The gamma matrices for the five dimensional case are four
dimensional matrices which satisfy the Clifford algebra
$\{\Gamma_{\mu},\Gamma_{\nu}\}=2 g_{\mu\nu}{\cal I}_{4 \times 4}$. A
representation for $\Gamma_{\mu}$ is obtained from the usual
representation $\gamma_{\mu}$ of the four dimensional QED, by
setting $\Gamma_{\mu}=\gamma_{\mu}$ for $\mu=0,1,2,3$ and
$\Gamma_{4}=i\gamma_{5}$.

It is well known that for odd dimensions, a parity \footnote{In the
case of five dimensions the symmetry of parity can be defined as
$x\rightarrow x$ and $x_{4}\rightarrow -x_{4}$. The fermionic field
is transformed as $\Psi(x,x_{4})\rightarrow
\Gamma_{4}\Psi(x,-x_{4})$, then the kinetic term of the Dirac
Lagrangian is invariant under the symmetry of parity, but the mass
term violates it.} violating term (Chern-Simons term\footnote{The
induced Chern-Simons term is of the form $\int d^{3}x \:
\varepsilon^{\mu\nu\rho\sigma\tau}A_{\mu}
\:\partial_{\nu}A_{\rho}\partial_{\sigma}A_{\tau}$, in the case of
five dimensions, and it is zero in the case we examine.}) is induced
by the quantum corrections. However this term is zero in the case of
the constant magnetic field.

Thus we will concentrate on the parity invariant term of the quantum
action which reads:
\begin{eqnarray}
S_{Q}[A]=\frac{1}{2i}Tr\ln(\not\!\!D^{2}+m^{2}_{f})
\end{eqnarray}

Instead of the effective action, it is more convenient for our
purposes to use the quantum part of the effective energy per unit of
the three dimensional volume $V=L^{3}$ (where $L$ is the size of the
three dimensional space box) which is given by the following
equation
\begin{equation}
E_{Q}[A]=-\frac{1}{VT}S_{Q}[A]
\end{equation}
where $T$ is the total length of time.

In this paper we aim to compute the effective action in the presence
of a constant magnetic field which is directed toward the $x_{3}$
axis. The vector potential that corresponds to this magnetic field
is $A=(0,0,B_{5}\; x_{1},0,0)$, where $B_{5}=F_{12}$ is the five
dimensional field strength. The magnetic field that corresponds to
four dimensions is $\vec{B}=B\vec{e}_{3}$ where $B=B_{5}\sqrt{2 \pi/
M}$. Also we define the four dimensional coupling constant $e$ as
$e=e_{5}\sqrt{M/2 \pi}$. Note that the product $eB(=e_{5}B_{5})$ is
independent of the dimensionality of space, and has dimension of
square of mass in natural units.

In what follows we will use the four dimensional quantities. For
example, the classical energy per unit of the three dimensional
volume can be expressed as
\begin{equation}
E_{class}=\frac{1}{2}\frac{2
\pi}{M}B_{5}^{2}=\frac{1}{2}B^{2}=\frac{1}{2e^{2}}(eB)^{2}
\end{equation}

From the integral representation
$\ln(a/b)=-\int_{0}^{+\infty}(ds/s)(e^{-as}-e^{-bs})$, if we use the
Eqs. (4) and (5) we obtain:
\begin{equation}
E_{Q}=-\frac{i}{2TV}\int_{0}^{+\infty}\frac{ds}{s}e^{-sm^{2}_{f}}\:
Tr(e^{-s\not D^{2}}-e^{-s\not
\partial^{2}})
\end{equation}
where we have renormalized by subtracting the effective energy for
zero magnetic field.

The trace in the above equation can be written as
\begin{equation}
Tr\:e^{-s\not
D^{2}}=Tr\:e^{-s(D^{2}+\frac{1}{2}e\Sigma^{\mu\nu}F_{\mu\nu})}=Tr\:e^{-sD^{2}}tr\:e^{-s\frac{1}{2}e\Sigma^{\mu\nu}F_{\mu\nu}}
\end{equation}
where $\Sigma^{\mu\nu}=\frac{i}{2}[\Gamma_{\mu},\Gamma_{\nu}]$.

By using the equation
\begin{equation}
e^{-s\frac{1}{2}e\Sigma^{\mu\nu}F_{\mu\nu}}=e^{-si\Gamma_{1}\Gamma_{2}eB}=\cosh(eBs){\cal
I}_{4\times4}-i\Gamma_{1}\Gamma_{2}\sinh(eBs)
\end{equation}
we obtain
\begin{equation}
tr\:e^{-s\frac{1}{2}e\Sigma^{\mu\nu}F_{\mu\nu}}=4\cosh(eBs)
\end{equation}
where we have used the identity $tr\Gamma_{1}\Gamma_{2}=0$.

The operator $D^{2}$ in the presence of the magnetic field is:
\begin{equation}
D^{2}=\partial_{0}^{2}-\partial_{1}^{2}-(\partial_{2}-ieBx_{1})^{2}-\partial_{3}^{2}-\partial_{4}^{2}
\end{equation}
We will be compute the trace $Tr\;e^{-s\;D^{2}}$ by using the
complete basis of eigenfunctions
\begin{equation}
 \Psi(x,x_{4})\sim e^{-i \omega
x_{0}} e^{ip_{2}x_{2}}e^{ip_{3}x_{3}} e^{i M m x_{4}}\: u(x_{1})
\quad (m=0,\pm 1,\pm 2,..)
\end{equation}

The function $u(x_{1})$ satisfies the eigenvalue equation
\begin{equation}
(-\partial_{1}^{2}+(p_{2}+eBx_{1})^{2})\:u(x_{1})=E_{n}\:u(x_{1})
\end{equation}
and the corresponding eigenvalues $E_{n}$ are the well known Landau
levels:
\begin{equation}
E_{n}=2eB(n+\frac{1}{2}) \quad (n=0,1,2,..)
\end{equation}
We remind the reader that in every Landau level corresponds an
infinite degeneracy factor with value $eB L^{2}/2 \pi$.

From Eqs. (11),(12),(13) and (14), if we perform a wick rotation
$\omega\rightarrow i\omega$, we obtain
\begin{eqnarray}
Tr\;e^{-s\;D^{2}}=i\frac{TL^{3}eB}{16 \pi^{2}}\frac{1}{s
\sinh(eBs)}\sum_{m=-\infty}^{+\infty}e^{-s\; M^{2}\:m^{2}}
\end{eqnarray}
where we have used the equation $\sum_{n}e^{-sE_{n}}=1/2\sinh(eBs)$.

By using Eqs. (10) and (15) we obtain
\begin{eqnarray}
E_{Q}=\frac{eB}{8
\pi^{2}}\int_{1/\Lambda^{2}}^{+\infty}\frac{ds}{s^{2}}\;e^{-s
\;m_{f}^{2}}\left(\coth(eBs)-\frac{1}{eBs}\right)F(s M^{2})
\end{eqnarray}
where we have set
\begin{eqnarray}
F(s)=\sum_{m=-\infty}^{+\infty}e^{-m^{2}s}
\end{eqnarray}
and we have rendered the integral of the Eq. (16) convergent by
introducing an ultraviolet cut-off \footnote{Note that the cut-off
$\Lambda$, which corresponds to the proper time-method, is not equal
to the physical cut-off scale $\Lambda_{Ph}$. However, we assume
that they are connected by a linear relation
$\Lambda_{Ph}=\sqrt{r}\Lambda$ where $r$ is of the order of unity. A
value for the parameter $r$ has been determined in Ref. \cite{12},
and it is found to be of the order of unity for all the values of
$\delta$. In addition, a different study for the parameter $r$, from
the point view of the Wilson renormalization group, can be found in
Ref. \cite{Z12}. However, in this work it is not necessary to
introduce the fundamental scale explicitly, and thus we will not
deal further with this topic.}$\Lambda$.

If we use the expansion
\begin{equation}
\coth(z)=\frac{1}{z}+\frac{z}{3}-\frac{z^{3}}{45}+O(z^{5})
\end{equation}
the Eq. (16) can be written as:
\begin{eqnarray}
E_{Q}&=&\frac{eB}{8
\pi^{2}}\int_{0}^{+\infty}\frac{ds}{s^{2}}\;e^{-s
\;m_{f}^{2}}\left(\coth(eBs)-\frac{1}{eBs}-\frac{eBs}{3}\right)F(s
M^{2})\nonumber \\&+&\frac{(eB)^{2}}{24
\pi^{2}}\int_{1/\Lambda^{2}}^{+\infty}\frac{ds}{s}\;e^{-s
\;m_{f}^{2}}F(s M^{2})
\end{eqnarray}

The last term, which is cut-off dependent, corresponds to the
diagram with one fermionic loop and two external legs \footnote{This
diagram is known as the vacuum polarization diagram, and the
external lengths represent the interaction of the fermion with the
classical magnetic field.}. This term can be incorporated in the
classical energy, as is shown bellow
\begin{eqnarray}
E_{eff}&=&E_{class}+E_{Q} \nonumber \\
&=&\frac{1}{2}\left(\frac{1}{e^{2}}+\frac{1}{12
\pi^{2}}\int_{1/\Lambda^{2}}^{+\infty}\frac{ds}{s}\;e^{-s
\;m_{f}^{2}}F(s M^{2}) \right)(eB)^{2} \nonumber
\\&+&\frac{eB}{8
\pi^{2}}\int_{0}^{+\infty}\frac{ds}{s^{2}}\;e^{-s
\;m_{f}^{2}}\left(\coth(eBs)-\frac{1}{eBs}-\frac{eBs}{3}\right)F(s
M^{2})
\end{eqnarray}
The renormalized coupling constant $e_{R}$ is defined as
\begin{eqnarray}
\frac{1}{e_{R}^{2}}=\frac{1}{e^{2}}+\frac{1}{12
\pi^{2}}\int_{1/\Lambda^{2}}^{+\infty}\frac{ds}{s}\;e^{-s
\;m_{f}^{2}}F(s M^{2})
\end{eqnarray}
Note that the coupling constant $e$ is the bare coupling constant
which corresponds to the scale $\Lambda_{Ph}$, and $e_{R}$ is the
renormalized coupling constant defined according to the first of the
renormalization conditions of Eq. (41) in section 4.2.

If we take that $eB$ is a renormalization group invariant quantity
(or $eB=e_{R}B_{R}$), we can write
\begin{eqnarray}
E_{eff}=\frac{1}{e_{R}^{2}}(e_{R}B_{R})^{2}+E^{(R)}_{Q}
\end{eqnarray}
where the renormalized quantum part of the effective energy
$E_{Q}^{(R)}$ is cut-off independent. If we set $z=eBs$ in Eq. (20)
we obtain
\begin{eqnarray}
E_{Q}&=&\frac{(eB)^{2}}{8
\pi^{2}}\int_{0}^{+\infty}\frac{dz}{z^{2}}\;e^{-\frac{z\;m_{f}^{2}}{eB}}\left(\coth(z)-\frac{1}{z}-\frac{z}{3}\right)F\left(\frac{z
\; M^{2}}{eB}\right)
\end{eqnarray}

For the sake of simplicity, we have dropped the index (R), otherwise
in the above equation we should write $e_{R}$, $B_{R}$ and
$E_{Q}^{(R)}$ .

We note that $F(zM^{2}/eB)\rightarrow 1$ as $eB/M^{2}<<1$. Thus the
Eq. (23) is reduced to the usual expression for the quantum energy
in the case of four dimensions, as is required.

\section{Strong field limit $(eB>>M^{2})$}

We remind the reader that the strong field asymptotic formula for
the case of four dimensions reads:
\begin{eqnarray}
E_{Q}^{asympt}=-\frac{(eB)^{2}}{24 \pi^{2}}\ln(\frac{eB}{m_{f}^{2}})
\end{eqnarray}
See for example Refs. \cite{15,18}.

Now if we use the Poisson formula we can prove that
\begin{equation}
F(s)=\sum_{m=-\infty}^{+\infty}e^{- m^{2}s}=\sqrt{\frac{\pi}{
s}}\sum_{r=-\infty}^{+\infty}e^{\frac{-\pi^{2}r^{2}}{s}}
\end{equation}
and from this equation we obtain $F(s)\sim \sqrt{\frac{\pi}{ s}}$ as
$s\rightarrow 0$.

In order to isolate the asymptotic behavior of Eq. (19) for
$s\rightarrow 0$, we define the function
\begin{equation}
L_{1}(s)=F(s)-\sqrt{\frac{\pi}{s}}
\end{equation}
which has the identity: $\lim_{s\rightarrow 0}s^{-n}L_{1}(s)=0$ for
every integer $n$. This asymptotic behavior for $s\rightarrow 0$ has
been confirmed numerically by plotting the function $s^{-n}L_{1}(s)$
for several values of $n$. Note that due to this identity the first
two integrals, in Eq. (27) below, are rendered convergent.

If we use the Eqs. (26) and (23) we obtain
\begin{eqnarray}
E_{Q}&=&\frac{eB}{8
\pi^{2}}\int_{0}^{+\infty}\frac{ds}{s^{2}}\;e^{-s
\;m_{f}^{2}}\left(\coth(eBs)-\frac{1}{eBs}\right)L_{1}(s
M^{2})-\frac{(eB)^{2}}{24
\pi^{2}}\int_{0}^{+\infty}\frac{ds}{s}\;e^{-s
\;m_{f}^{2}} L_{1}(s M^{2})\nonumber \\
&+&\frac{eB}{8
\pi^{3/2}M}\int_{0}^{+\infty}\frac{ds}{s^{5/2}}\;e^{-s
\;m_{f}^{2}}\left(\coth(eBs)-\frac{1}{eBs}-\frac{eBs}{3}\right)
\end{eqnarray}

In order to study the asymptotic behavior of $E_{Q}$ in the strong
magnetic field limit $eB>>M^{2}$ we will study separately the three
integrals that appear in Eq. (27). It will be convenient to call
them $E^{(1)}_{Q}$, $E^{(2)}_{Q}$ and $E^{(3)}_{Q}$ respectively.

If we set $y=s M^{2}$ in the first integral of Eq. (27) we obtain
\begin{eqnarray}
E^{(1)}_{Q}&=&\frac{{(eB)}^{2}}{8 \pi^{2}}\left(\frac{eB
}{M^{2}}\right)^{-1}\int_{0}^{+\infty}\frac{dy}{y^{2}}\;e^{-y
\;\frac{m_{f}^{2}}{M^{2}}}\left(\coth(\frac{eBy}{M^{2}})-\frac{1}{eBy/M^{2}}\right)
L_{1}(y)
\end{eqnarray}
and for $\frac{eB}{M^{2}}\rightarrow +\infty$ we find
\begin{eqnarray}
E^{(1)}_{Q}&\rightarrow&\frac{(eB)^{2}}{8 \pi^{2}}\left(\frac{eB
}{M^{2}}\right)^{-1}\int_{0}^{+\infty}\frac{dy}{y^{2}}\;e^{-y
\;\frac{m_{f}^{2}}{M^{2}}} L_{1}(y)
\end{eqnarray}

Similarly if we set $z=eBs$ in the third integral of Eq. (27) we
obtain
\begin{eqnarray}
E^{(3)}_{Q}=\frac{(eB)^{2}}{8
\pi^{3/2}}\sqrt{\frac{eB}{M^{2}}}\int_{0}^{+\infty}\frac{dz}{z^{5/2}}\;e^{-\frac{z
\:m_{f}^{2}}{eB}} \left(\coth(z)-\frac{1}{z}-\frac{z}{3}\right)
\end{eqnarray}
and for $eB>>M^{2}>>m_{f}^{2}$ we see that
\begin{eqnarray}
E_{Q}^{(3)}\rightarrow\frac{(eB)^{2}}{8
\pi^{3/2}}\sqrt{\frac{eB}{M^{2}}}\int_{0}^{+\infty}\frac{dz}{z^{5/2}}
\left(\coth(z)-\frac{1}{z}-\frac{z}{3}\right)=-\frac{c_{1}(eB)^{2}}{8
\pi^{3/2}}\sqrt{\frac{eB}{M^{2}}}
\end{eqnarray}
where $c_{1}$ is a positive constant. For the numerical value of
$c_{1}$ see Table \ref{delta}.

The integral
$$E_{Q}^{(2)}=-\frac{(eB)^{2}}{24
\pi^{2}}\int_{0}^{+\infty}\frac{ds}{s}\;e^{-s \;m_{f}^{2}} L_{1}(s
M^{2})$$ is obviously proportional to $(eB)^{2}$ and no further
analysis is needed.

If we compare the equations (28) and (30) we see that $E_{Q}^{(3)}$
dominates in the strong field limit. Thus, the leading term of the
asymptotic formula for the quantum energy $E_{Q}$ is
\begin{eqnarray}
E_{Q}^{(asympt)}=-\frac{c_{1}(eB)^{2}}{8
\pi^{3/2}}\sqrt{\frac{eB}{M^{2}}}
\end{eqnarray}
and the next to leading term is given by the integral $E_{Q}^{(2)}$.

Comparing Eq. (32) with Eq. (24) we see that the logarithmic
dependence for four dimensions has been modified with a square root
law in the case of five dimensions.

\section{More than five dimensions (D=4+$\delta$)}

In this section we aim to generalize the asymptotic formula of Eq.
(32) for more than five dimensions. We set $D=\delta+4$ where
$\delta$ is the number of the extra compact dimensions. However, if
we assume that the number of extra dimensions is restricted by the
string theory, it can not exceed the number six (or $1\leq\delta\leq
6$). Also we assume that the radius $R=1/M$ is the same for all the
extra dimensions.

An obvious modification, in order to extend Eq. (16) for a general
number of extra dimensions $\delta$, is to do the replacement
$F(sM^{2})\rightarrow \left(F(sM^{2})\right)^{\delta}$. A second
modification is to multiply Eq. (16) by $2^{[D/2]-2}$, which is due
to the trace of gamma matrices \footnote{We remind the reader that
the gamma matrices, in a D-dimensional space, have dimensions
$2^{[D/2]}\times2^{[D/2]}$.}
$tr\:e^{-s\frac{1}{2}e\Sigma^{\mu\nu}F_{\mu\nu}}=2^{[D/2]}\cosh(eBs)$.

Now it is a straightforward matter to generalize the formula of Eq.
(16) in the case of $\delta$ extra dimensions
\begin{eqnarray}
E_{Q}=2^{[D/2]-2} \frac{eB}{8
\pi^{2}}\int_{1/\Lambda^{2}}^{+\infty}\frac{ds}{s^{2}}\;e^{-s
\;m_{f}^{2}}\left(\coth(eBs)-\frac{1}{eBs}\right)\left(F(s
M^{2})\right)^{\delta}
\end{eqnarray}
where $e=e_{D}(M/2 \pi)^{\delta/2}$, $B=B_{D}(2\pi/M)^{\delta/2}$ ,
and $e_{D}$ and $B_{D}$ are the corresponding D dimensional
quantities.

Note that for $eB/M^{2}<<1$ the above equation is not reduced to the
corresponding four dimensional expression, as is required. In
particular it differs by a factor $2^{[D/2]-2}$. This is due to the
fact that the reduced theory contains $2^{[D/2]-2}$ four dimensional
Dirac spinors with the same mass term, and not one as happens with
the four dimensional model. Thus we can make contact with the four
dimensional quantum energy by dividing with the number of Dirac
spinors $2^{[D/2]-2}$. A more sophisticated way to solve this
problem would be an orbifold model with the appropriate boundary
conditions, so that only one Dirac spinor would survive in four
dimensions. However in this work we will not perform computations
for this case.

An interesting point is that the strong field ($eB>>M^{2}$)
asymptotic behavior for the effective energy, is independent from
the details of the compactification. The reason is that, for
$eB>>M^{2}$, the extra dimensions lose their compact structure and
behave as if they were noncompact. Thus, even if we had assumed
another compactification scenario, for example an orbifold model, we
would obtain exactly the same result for the effective energy in the
strong field limit.

The next step is to subtract and to add back the vacuum
polarization diagram, which is responsible for the renormalization
of the coupling constant,
\begin{eqnarray}
 E_{Q}&=&2^{[D/2]} \frac{eB}{32
\pi^{2}}\int_{1/\Lambda^{2}}^{+\infty}\frac{ds}{s^{2}}\;e^{-s
\;m_{f}^{2}}\left(\coth(eBs)-\frac{1}{eBs}-\frac{eBs}{3}\right)\left(F(s
M^{2})\right)^{\delta}\nonumber \\&+&2^{[D/2]} \frac{(eB)^{2}}{96
\pi^{2}}\int_{1/\Lambda^{2}}^{+\infty}\frac{ds}{s}\;e^{-s
\;m_{f}^{2}} \left(F(s M^{2})\right)^{\delta}
\end{eqnarray}
In addition, the renormalized coupling constant is
\begin{eqnarray}
\frac{1}{e_{R}^{2}}=\frac{1}{e^{2}}+\frac{2^{[D/2]}}{48
\pi^{2}}\int_{1/\Lambda^{2}}^{+\infty}\frac{ds}{s}\;e^{-s
\;m_{f}^{2}}\left(F(s M^{2})\right)^{\delta}
\end{eqnarray}

\subsection{Effective Energy for $\delta=1,2,3$}

For $\delta=1,2,3$, the renormalized quantum energy reads:
\begin{eqnarray}
 E_{Q}&=&2^{[D/2]} \frac{(eB)^{2}}{32
\pi^{2}}\int_{0}^{+\infty}\frac{dz}{z^{2}}\;e^{-\frac{z
\;m_{f}^{2}}{eB}}\left(\coth(z)-\frac{1}{z}-\frac{z}{3}\right)\left(F\left(\frac{z
\;M^{2}}{eB}\right)\right)^{\delta}
\end{eqnarray}
Note that the above integral is convergent only for $\delta=1,2,3$,
and the quantities e and B that appear in this integral are the
renormalized ones, for which we have dropped the index (R).

The strong field limit is obtained by using exactly the same
method, which was presented in the previous section. The
corresponding asymptotic formula reads:
 \begin{eqnarray}
E_{Q}^{asympt}=-c_{\delta}\frac{2^{[D/2]}(eB)^{2}}{32\;
\pi^{(4-\delta)/2}}\left(\frac{eB}{M^{2}}\right)^{\delta/2} \qquad
(\delta=1,2,3)
\end{eqnarray}
where
\begin{eqnarray}
c_{\delta}=-\int_{0}^{+\infty}\frac{dz}{z^{(\delta+4)/2}}
\left(\coth(z)-\frac{1}{z}-\frac{z}{3}\right) \quad (\delta=1,2,3)
\end{eqnarray}
The numerical values of the constant $c_{\delta}>0$ are given in
Table \ref{delta}.

In order to prove the asymptotic formula of Eq. (37) we have used
the function
\begin{equation}
L_{\delta}(s)=\left(F(s)\right)^{\delta}-\left(\frac{\pi}{
s}\right)^{\frac{\delta}{2}}
\end{equation}
Note that the identity $\lim_{s\rightarrow 0}s^{-n}L_{\delta}(s)=0$
is valid also in the case of more than one extra dimensions, and
that it has been confirmed numerically.

\subsection{Effective Energy for $\delta=4,5,6$}

For $\delta=4,5,6$ the integral in Eq. (36) is divergent for $s=0$.
Thus, the cut-off $\Lambda$ appears explicitly in the expression for
the quantum energy. It is possible to isolate the cut-off dependent
part by subtracting and adding back the Feynman diagram with one
fermion loop and four external legs, or
\begin{eqnarray}
E_{Q}&=& 2^{[D/2]} \frac{eB}{32
\pi^{2}}\int_{0}^{+\infty}\frac{ds}{s^{2}}\;e^{-s
\;m_{f}^{2}}\left(\coth(eBs)-\frac{1}{eBs}-\frac{eBs}{3}\right) \nonumber \\
&+&2^{[D/2]} \frac{eB}{32
\pi^{2}}\int_{0}^{+\infty}\frac{ds}{s^{2}}\;e^{-s
\;m_{f}^{2}}\left(\coth(eBs)-\frac{1}{eBs}-\frac{eBs}{3}+\frac{(eBs)^{3}}{45}\right)\left(\left(F(s
M^{2})\right)^{\delta}-1\right)
 \nonumber \\ &-&2^{[D/2]}
\frac{(eB)^{4}}{1440
\pi^{2}}\int_{1/\Lambda^{2}}^{+\infty}ds\;s\;e^{-s
\;m_{f}^{2}}\left(\left(F(s M^{2})\right)^{\delta}-1\right)
\end{eqnarray}
In this case there is not a classical term in which the cut-off
dependent term of the above equation can be incorporated. Of course,
this is a problem that is due to the nonrenormalizable character of
the model we examine. However, we will overcome this problem by
assuming the existence of a higher dimension operator of the form
$w_{D}/2\int d^{D}x (F_{\mu\nu}F^{\mu\nu})^{2}$. The cut-off
dependent term in Eq. (40) can be incorporated in this higher
dimension operator. The renormalization conditions (see also Ref.
\cite{Savvidy:1}) according to which we separate the cut-off
dependent part from the finite part are
\begin{equation}
\frac{1}{e^{2}}\left[\frac{d E_{eff}}{d {\cal F}}\right] _{{\cal
F}=0}=\frac{1}{e^{2}_{R}}, \qquad \frac{1}{4}\left[\frac{d^{2}
E_{eff}}{d {\cal F}^{2}}\right] _{{\cal F}=0}=w_{R}-
\frac{2^{[D/2]}}{1440 \pi^{2}}\frac{e^{4}}{m_{f}^{4}}
\end{equation}
where ${\cal F}=\frac{1}{4}F_{\mu\nu}F^{\mu\nu}$, and $w_{R}$ is the
corresponding renormalized four dimensional coupling constant. Thus,
the effective energy includes a term of the form $w_{R}/2 \;B^{4}$
where $w_{R}$ is a free parameter, which can be determined
experimentally according to the renormalization conditions of Eq.
(41). Note, that we have included the term
$-\frac{2^{[D/2]}}{1440\pi^{2}}\frac{e^{4}}{m_{f}^{4}}$ in Eq. (40),
in order to make contact with the four dimensional result for the
quantum energy in the weak field limit $eB<<M^{2}$.

The renormalized quantum energy consists of the first two terms of
Eq. (40):
\begin{eqnarray}
E^{(R)}_{Q}&=& 2^{[D/2]} \frac{eB}{32
\pi^{2}}\int_{0}^{+\infty}\frac{ds}{s^{2}}\;e^{-s
\;m_{f}^{2}}\left(\coth(eBs)-\frac{1}{eBs}-\frac{eBs}{3}\right)\\
&+&2^{[D/2]} \frac{eB}{32
\pi^{2}}\int_{0}^{+\infty}\frac{ds}{s^{2}}\;e^{-s
\;m_{f}^{2}}\left(\coth(eBs)-\frac{1}{eBs}-\frac{eBs}{3}+\frac{(eBs)^{3}}{45}\right)\left(\left(F(s
M^{2})\right)^{\delta}-1\right) \nonumber
\end{eqnarray}
The corresponding strong field asymptotic formula for $\delta=5,6$
is
\begin{eqnarray}
E_{Q}^{asympt} =\frac{c_{\delta}\;2^{[D/2]}(eB)^{2}}{32\;
\pi^{(4-\delta)/2}}\left(\frac{eB}{M^{2}}\right)^{\delta/2} \quad
(\delta=5,6)
\end{eqnarray}
where
\begin{eqnarray}
c_{\delta}=\int_{0}^{+\infty}\frac{dz}{z^{(4+\delta)/2}}\left(\coth(z)-\frac{1}{z}-\frac{z}{3}+\frac{{z}^{3}}{45}\right)\quad
(\delta=5,6)
\end{eqnarray}
Note, that from the first term of Eq. (42) we obtain the logarithmic
asymptotic behavior of Eq. (24), which corresponds to the case of
four dimensions.

For $\delta=4$ we can show that the strong field asymptotic formula
reads:
\begin{eqnarray}
E_{Q}^{asympt} =\frac{2^{[D/2]}(eB)^{2}}{1440}
\left(\frac{eB}{M^{2}}\right)^{\delta/2} \ln(\frac{eB}{M^{2}})\quad
(\delta=4)
\end{eqnarray}
The logarithmic factor $\ln(eB/M^{2})$ in the above asymptotic
formula, for $\delta=4$, is due to the logarithmic divergent of the
second integral of Eq. (42), as $eB/M^{2}\rightarrow +\infty$.

Note, that for $\delta=4,5,6$ the asymptotic formula is positive,
contrary to the case of $\delta=1,2,3$.
\begin{table}
  \centering
\begin{tabular}{|c|c|c|c|c|c|c|}
  \hline
  $\delta$ & 1 &2 & 3 & 4 & 5 & 6 \\
  \hline
  $c_{\delta}$ & 0.340 &0.122  &0.087 & - &0.027 &0.011 \\
  \hline
   $f_{\delta}$ & 0.219 &0.091  &0.074 & - &0.024 &0.010 \\
  \hline
\end{tabular}
  \caption{The constants $c_{\delta}$ and $f_{\delta}$ for $\delta=1,2,3,5,6$.}\label{delta}
\end{table}

\section{Quantum energy for scalar fields}

In this section we will discus briefly the quantum energy for a U(1)
gauge field theory with scalar fields. The path integral in this
case is
\begin{equation}
Z=\int_{b.c} {\cal D}A \:{\cal D}\phi^{*}\: {\cal D}\phi \:e^{i\int
d^{D}x
(-\frac{1}{4}F_{\mu\nu}F^{\mu\nu}+(D^{\mu}\phi)^{*}(D_{\mu}\phi)-m_{s}^{2}
\phi^{*}\phi)}
\end{equation}
with boundary conditions $\phi(x,x_{4})=\phi(x,x_{4}+2 \pi R)$.

The quantum part of the effective action is
\begin{eqnarray}
S_{Q}[A]=\frac{1}{i}\ln\left(\int_{b.c}{\cal D}\phi^{*} {\cal D}\phi
\; e^{-i\int d^{D} x\;
\phi^{*}(D^{\mu}D_{\mu}+m_{s}^{2})\phi}\right)=-\frac{1}{i}\;Tr\ln(D^{2}+m_{s}^{2})
\end{eqnarray}

In a similar way with that of previous sections, we find that the
renormalized quantum energy in the case scalar fields is
\begin{eqnarray}
E_{Q}&=&-\frac{(eB)^{2}}{16
\pi^{2}}\int_{0}^{+\infty}\frac{dz}{z^{2}}\;e^{-\frac{z
\;m_{s}^{2}}{eB}}\left(\frac{1}{\sinh(z)}-\frac{1}{z}+\frac{z}{6}\right)\left(F\left(\frac{zM^{2}}{eB}\right)\right)^{\delta}
\end{eqnarray}
for $\delta=1,2,3$.

The corresponding strong field asymptotic formula for the quantum
energy is
\begin{equation}
E_{Q}^{asympt}=-f_{\delta}\frac{(eB)^{2}}{16\;\pi^{(4-\delta)/2}}\left(\frac{eB}{M^{2}}\right)^{\delta/2}\quad
(\delta=1,2,3)
\end{equation}
where $f_{\delta}$ is a positive constant which is given by the
equation
\begin{equation}
f_{\delta}=\int_{0}^{+\infty}\frac{dz}{z^{(\delta+4)/2}}\left(\frac{1}{\sinh(z)}-\frac{1}{z}+\frac{z}{6}\right)
\quad (\delta=1,2,3)
\end{equation}

For $\delta=5,6$ we obtain
\begin{eqnarray}
E_{Q}^{asympt} =f_{\delta}\frac{(eB)^{2}}{16\;
\pi^{(4-\delta)/2}}\left(\frac{eB}{M^{2}}\right)^{\delta/2} \quad
(\delta=5,6)
\end{eqnarray}
where $f_{\delta}$ is positive, and it is given by the equation
\begin{eqnarray}
f_{\delta}=-\int_{0}^{+\infty}\frac{dz}{z^{(4+\delta)/2}}\left(\frac{1}{\sinh(z)}-\frac{1}{z}+\frac{z}{6}-\frac{7
{z}^{3}}{360}\right)\quad (\delta=5,6)
\end{eqnarray}
For the numerical values of the constants $f_{\delta}$ see table
\ref{delta}.

Finally, for $\delta=4$ we obtain
\begin{eqnarray}
E_{Q}^{asympt} =\frac{7
(eB)^{2}}{5760}\left(\frac{eB}{M^{2}}\right)^{\delta/2}
\ln(\frac{eB}{M^{2}})\quad (\delta=4)
\end{eqnarray}

\section{Numerical results for the quantum energy}

\begin{figure}[h]
\begin{center}
\includegraphics[scale=1,angle=0]{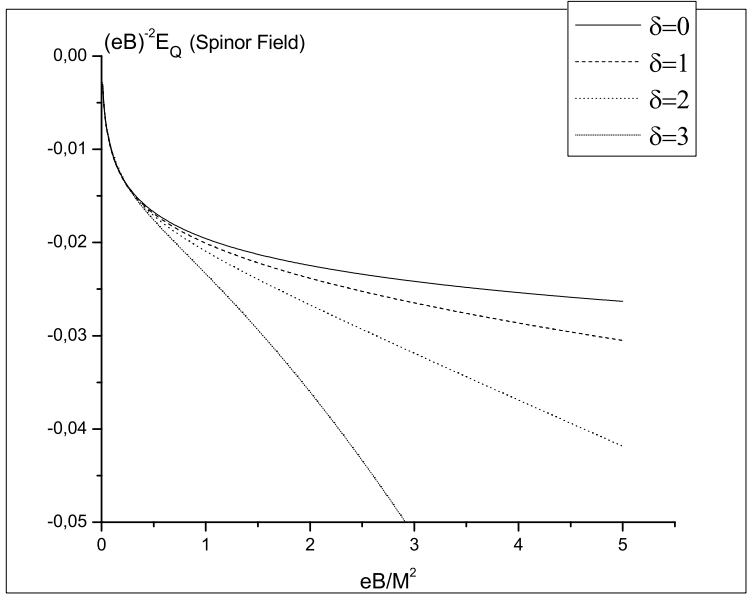}
\end{center}
\caption {$(2^{[D/2]-2})^{-1}(eB)^{-2}E_{Q}$ versus $eB/M^{2}$, for
fixed ratio $M/m_{f}=10^{3}$, and $\delta=0,1,2,3$, for the fermion
field.} \label{gr1}
\end{figure}

\begin{figure}[h]
\begin{center}
\includegraphics[scale=1,angle=0]{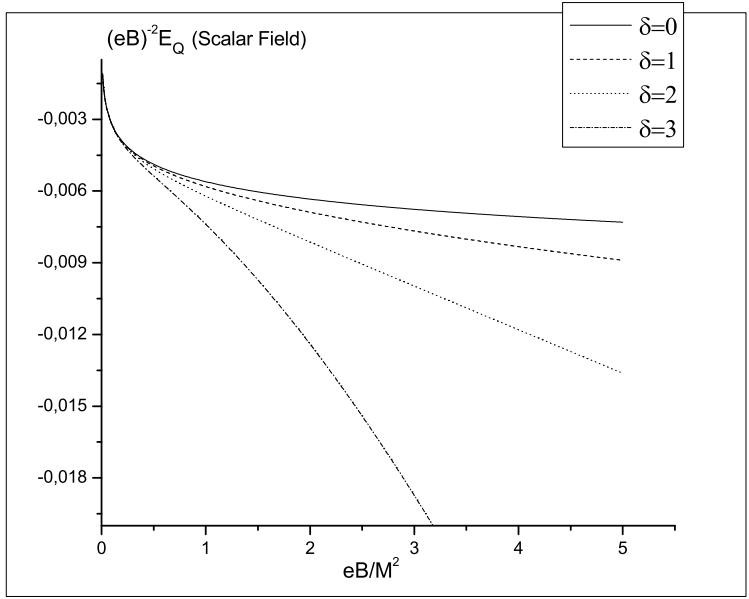}
\end{center}
\caption {$(eB)^{-2}E_{Q}$ versus $eB/M^{2}$, for fixed ratio
$M/m_{s}=10^{3}$, and $\delta=0,1,2,3$, for the scalar field.}
\label{gr2}
\end{figure}

 In Fig. \ref{gr1} we have plotted the
 $(2^{[D/2]-2})^{-1}(eB)^{-2}E_{Q}$ as a function of $eB/M^{2}$, in the case of fermions.
 This figure confirms numerically
 the strong field asymptotic behavior for the quantum energy,
 for $\delta=1,2,3$ (see Eq. (37)). In addition, for $eB<<M^{2}$ the quantum energy is
 independent from the extra dimensions and it coincides with the four dimensional result, as
 is required. Thus the existence of the extra dimensions is not
 observable for weak magnetic fields $eB<<M^{2}$. The corresponding case for the scalar
 fields is presented in Fig. \ref{gr2}, and as we see, it has the same features with that
 of the fermion fields (see also Eq. (49)).

\begin{figure}[h]
\begin{center}
\includegraphics[scale=1,angle=0]{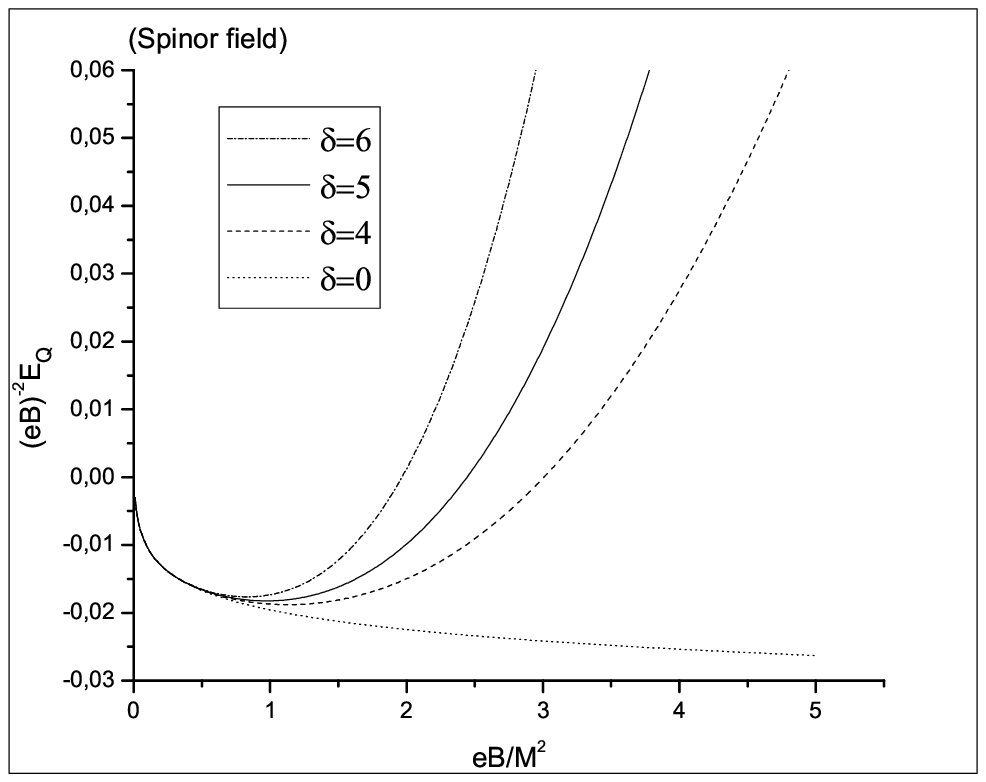}
\end{center}
\caption {$(2^{[D/2]-2})^{-1}(eB)^{-2}E_{Q}$ versus $eB/M^{2}$, for
fixed ratio $M/m_{f}=10^{3}$, $\delta=0,4,5,6$, for the fermion
field.} \label{gr3}
\end{figure}

\begin{figure}[h]
\begin{center}
\includegraphics[scale=1,angle=0]{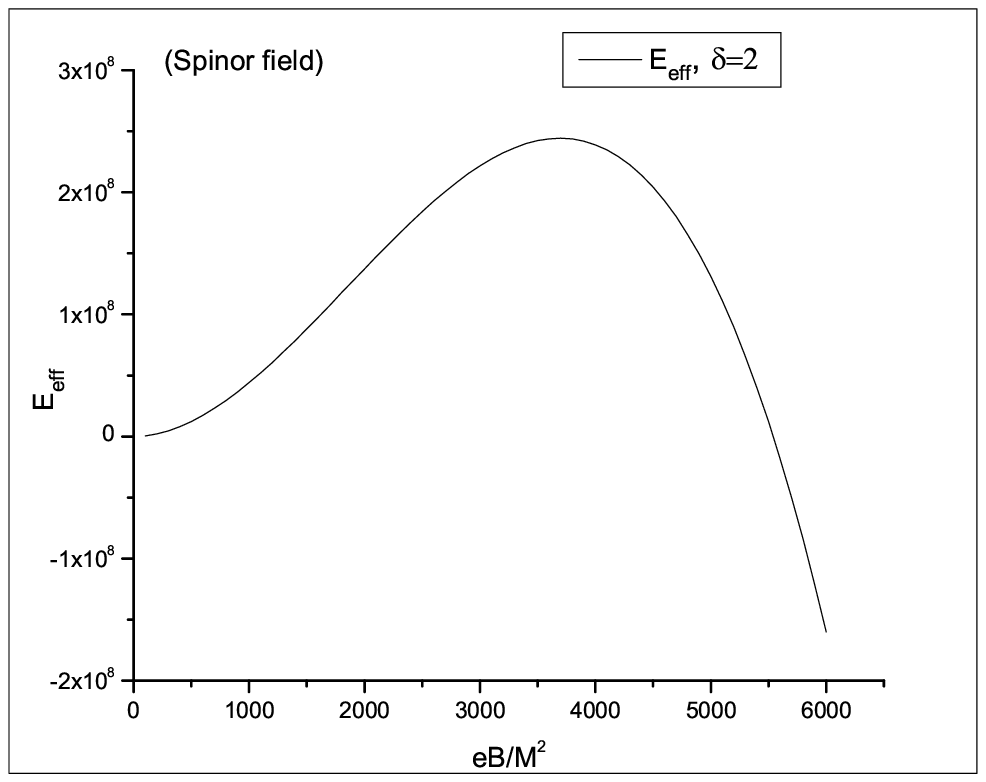}
\end{center}
\caption {$E_{eff}=E_{class}+E_{Q}$ versus $eB/M^{2}$, for fixed
ratio $M/m_{f}=2 10^{6}$, $\delta=2$ and $e^{2}/4 \pi=1/137$ for the
fermion field.} \label{gr4}
\end{figure}

In Fig. \ref{gr3} we have plotted $(2^{[D/2]-2})^{-1}(eB)^{-2}E_{Q}$
as a function of $eB/M^{2}$ for the spinor field, in the case of
$\delta=0,4,5,6$. From the figure we observe that the extra
dimensions does not alter the four dimensional quantum energy in the
limit $eB<<M^{2}$, as is required. Note that for $eB>>M^{2}$  the
quantum energy for $\delta=4,5,6$ is positive, contrary to the case
of $\delta=1,2,3$. The corresponding figure, in the case of the
scalar fields, has the same features with those of Fig. \ref{gr3}.

Finally, in Fig. \ref{gr4}, we present a typical plot of the
effective energy $E_{eff}=E_{class}+E_{Q}$ as a function of
$eB/M^{2}$, in the case of the fermion fields. The corresponding
figure in the case of the scalar fields is not presented, as it
exhibits exactly the same features with those of Fig. \ref{gr4}. Of
course, Fig. \ref{gr4} is not reliable for large values of the
magnetic field, at which the perturbation theory breaks down (or
$|E_{Q}|=E_{class}$). In addition, we remind the reader that the
contributions of the possible higher dimensional operators has been
completely ignored.

For the numerical computation we used the Eq. (36) for
$\delta=1,2,3$ and the Eq. (42) for $\delta=4,5,6$.

We would like to estimate for which magnetic fields the quantum
energy becomes comparable with the quantum energy. In Table
\ref{comp} we determine the magnetic fields for which $|E_{Q}|=0.1
E_{class}$ and $|E_{Q}|=E_{class}$.

\begin{table}
  \centering
\begin{tabular}{|c|c|c|c|c|c|}
  \hline
  $|E_{Q}|=$& M&$\delta$=0 &$\delta$=1 &$\delta$=2 &$\delta$ =3  \\
  \hline
  $0.1 E_{class}$&1 Tev &$B\sim10^{70} G$ &$B\sim 8.6\; 10^{29}G$ &$B\sim9.5\;10^{27}G$ & $B\sim2.1\; 10^{27}G$  \\
    \hline
   $E_{class}$& 1 Tev &$B\sim 10^{574}G$ &$B\sim 8.6 \;10^{31}G$  &$B\sim9.5 \;10^{28}G$ & $B\sim9.5\;10^{27}G$  \\
  \hline
   $0.1 E_{class}$& 20 Tev &$B\sim10^{70} G$ &$B\sim3.5\;10^{32}G$ &$B\sim3.8\; 10^{30}G$ & $B\sim 8.5\; 10^{29}G$  \\
    \hline
   $E_{class}$& 20 Tev &$B\sim 10^{574}G$ &$B\sim 3.5 \;10^{34}G$  &$B\sim 3.8\;10^{31}G$ & $B\sim 3.9 \;10^{30}G$  \\
  \hline
\end{tabular}
  \caption{We present a comparison between the quantum and the classical energy, and in particular we
  determine the values of the magnetic fields for which $|E_{Q}|=0.1 E_{class}$
   and  $|E_{Q}|=E_{class}$ for $\delta=0,1,2,3$, $M=1Tev, 20Tev$, $e^{2}/4 \pi=1/137$, and $m_{f}=0.5 Mev$. We have not included the
   cases of $\delta=4,5,6$, as the value of $w_{R}$ is not known.}\label{comp}
  \end{table}

In the case of four dimensions, the strong field
($B>>e/m_{f}^{2}\sim 10^{14}G$ for $m_{f}=0.5 Mev$ and $e^{2}/4
\pi=1/137$) asymptotic behavior of the quantum energy is given by
the formula
$E_{Q}=-\frac{(eB)^{2}}{24\pi^{2}}\ln(\frac{eB}{m_{f}^{2}})$ (see
Eq. (24)). The quantum energy is equal to the ten per cent of the
classical energy (or $|E_{Q}|=0.1 E_{class}$) for magnetic fields of
the order $B\sim10^{70} G$. The value of the magnetic field for
which the perturbation theory breaks down (or $|E_{Q}|=E_{class}$)
is $B\sim10^{574} G$. These values are entirely unrealistic. Of
course, this is due to the small coupling constant of QED, and the
weak logarithmic growth of the quantum energy.

In the case of extra dimensions, and in particular for $\delta=3$
(for $\delta=1,2$ see Table \ref{comp}), we see that the quantum
energy becomes equal to the ten per cent of the classical energy for
$B\sim 10^{27} G$ $(M=1 Tev)$, and the perturbation theory breaks
down for $B\sim 10^{28} G$. These values are significantly smaller
than the corresponding values of the four dimensional case, but they
are still very large. For example the magnetic fields in neutron
stars are of the order of $B\sim 10^{14}G$, and the magnetic fields
which might have been created during the electroweak phase
transition are of the order of $B\sim 10^{24}G$ \cite{19}, which are
smaller than $B\sim 10^{27} G$ where the effects of the extra
dimensions become significant. However, the authors of Ref.
\cite{17}, have assumed the existence of magnetic fields
$B\sim10^{33}G$, in the early universe, in order to explain the
present-day galactic magnetic fields. Thus, for magnetic fields of
this order of magnitude the effects of the extra dimensions should
be taken into account (see also the discussion in the next section).

\section{Conclusions and Discussion}

We considered a possible generalization of a U(1) gauge field theory
with fermion fields (or scalar fields) in the case of $\delta$ extra
compact dimensions, and we presented a computation for the fermion
induced quantum energy, in the presence of a constant magnetic field
which is directed towards the $x_{3}$ axis.

Moreover, we would like to note that realist magnetic fields can not
cover an infinite volume. However, in the strong field limit the
typical length scale $1/\sqrt{eB}$ of the magnetic field is much
smaller than the spatial size $L$, or $eB \;L^{2}>>1$, and in this
case the magnetic field behaves as if it was constant everywhere.
Thus, we expect the formulas we derive to be valid, even if the
magnetic field vanishes outside a large volume $V=L^{3}$.

The model we examined is nonrenormalizable, and it should be assumed
as a low energy effective theory with a large physical cut-off
$\Lambda_{Ph}$ (Wilsonian effective field theory). A remarkable
feature is that the one loop result for the quantum energy is
independent of the cut-off $\Lambda_{Ph}$ for small $\delta$ (or
$\delta=1,2,3$). However, for $\delta=4,5,6$ the nonrenormalizable
character of the model worsens and the scale $\Lambda_{Ph}$ (or the
proper time cut-off $\Lambda$) appears explicitly in the quantum
energy. In this case we included a higher dimension operator of the
form $w_{D}/2\int d^{D}x (F_{\mu\nu}F^{\mu\nu})^{2}$ in the
Lagrangian of the model. Then, the cut-off dependent part of Eq.
(40) can be incorporated in this term according to the
renormalization conditions of Eq. (41). Thus, in the case of
$\delta=4,5,6$,the effective energy contains a term of the form
$\frac{w_{R}}{2} B^{4}$, with an additional free parameter $w_{R}$.

We would like to note that there are alternate ways to treat this
difficulty. For example, we can ignore all the higher dimension
operators and assume that the proper time cut-off $\Lambda$ is not
just a typical regularization parameter, but it is connected with
the physical cut-off $\Lambda_{Ph}$ via a linear relation (see Ref.
\cite{12}). Then, the result for the effective energy, for
$\delta=4,5,6$, will be a finite quantity which depends on the
physical cut-off $\Lambda_{Ph}$ (see Eq. (40)). However, in this
paper, we adopt the philosophy which is presented in Ref.
\cite{Wilson:1} and we incorporate the cut-off dependent term in a
higher dimension operator, as we explain in the previous paragraph
and in section 4.

We also studied the effect of the extra dimensions on the strong
field ($eB>>M^{2}$) asymptotic behavior of the effective energy. We
see that there is a critical value of the magnetic field
$B_{cr}=M^{2}/e$ ($\sim 10^{26} G$ for M=1 Tev), above which the
extra dimensions behave as if they were noncompact (see the figures
in section 6), and as a consequence the quantum energy increases
rapidly according to a power law behavior (see Eqs. (37),(43) and
(45)). This behavior modifies the weak logarithmic growth of the
quantum energy in the case of four dimensions (see Eq. (24)).

In this work we have not examined the interesting case of massless
fermions. We preferred to assume that the fermions are massive in
order to make contact with the Schwinger formula for four
dimensions, which has been formulated only for massive fermions.
However, our results for the strong field asymptotic behavior are
valid even for massless fermions. In this case the mass of the
fermions $m_{f}$ is viewed as an infrared cut-off parameter. An
interesting feature of the strong field asymptotic formulas, of Eq.
(32), (37), (43) and (45), is that they are independent of the
infrared cut-off $m_{f}$, for $eB>>M^{2}>>m_{f}^{2}$.

It is interesting to note that in higher dimensional spaces constant
magnetic fields are characterized by the electromagnetic tensor
$F_{\mu\nu}$ where $\mu,\nu=1,2,..D-1$, which has more than one
independent components (see for example Ref. \cite{extra1}). These
more general magnetic fields are physical objects which are
introduced by the higher dimensional theory. Even these fields have
not been observed directly in low energy physics, they can be
relevant physically indirectly. For example, in Ref. \cite{mag:sup},
the existence of a magnetic field in the intrinsic space can be used
as a tool for supersymmetry breaking. In this paper, we have assumed
the special case of a magnetic field which is directed toward the
$x_{3}$ axis of the physical space. The computation of the quantum
energy in the most general case of a higher dimensional magnetic
field is beyond the scope of this paper and could be a topic of
further investigation.

We emphasize again, that this extra-dimension U(1) model with
massive fermions (or with a massive scalar field), can not be viewed
as an extra-dimension extension of QED. It is only a toy model for
the study of the effects of the extra dimensions. Thus, the results
of this paper are only suggestive and not realistic.

However, we can use the $U(1)$ model in order to extract results for
a more realistic case. We will assume an extra-dimension version of
the standard model, before the electroweak symmetry
($SU_{L}(2)\times U_{Y}(1)$) breaking. Then the $U(1)$ symmetry, of
the $U(1)$ model of this work, corresponds to the $U_{Y}(1)$ of the
standard model, and the magnetic field corresponds to a hypercharge
$U_{Y}(1)$ magnetic field. Then the effective energy, per unit
volume, for $\delta=4,5,6$, in the case of $U_{Y}(1)$ reads:
\begin{eqnarray}
E_{eff}=\frac{1}{2g^{2}_{R}} (g_{R}B)^{2}+ \frac{w_{R}}{2g^{4}_{R}}
(g_{R}B)^{4}+E_{Q} \nonumber
\end{eqnarray}
where $g_{R}$ is the renormalized hypercharge $U_{Y}(1)$ coupling,
and $w_{R}$ is a free parameter (see section 4). The contributions
to quantum energy are come from the spinor and scalar fields of the
extra-dimension version of the standard model, and not from the
vector fields. If we take into account that the main features of the
quantum energy are the same for spinor and scalar fields (see
sections 4 and 5), we expect the quantum energy $E_{Q}$, for
$g_{R}B>>M^{2}$ and $\delta=4,5,6$, to be positive and to behave as
$E^{asympt}_{eff}\sim (eB)^{2} (eB/M)^{\delta/2}$ (for $\delta=4$
see Eq. (45)). Thus, for an appropriate negative value of the free
parameter $w_{R}$ it is possible for the effective energy to exhibit
a minimum for $g_{R} B>>M^{2}$, which corresponds to a stable
hypercharge $U_{Y}(1)$ magnetic field. Note, that in the above
mechanism we have completely ignore the effects of the other
possible higher order operators, and the negative value of the free
parameter $w_{R}$ has been put by hand.

It is worth to note that there are several scenarios for the
generation of primordial magnetic fields, see for example Ref.
\cite{prim:1}. Especially in the case of string cosmology,
sufficiently large seeds for generating the observed galactic
magnetic fields can be obtained from the amplification of
electromagnetic vacuum fluctuations, due to the inflationary
dynamics of the dilaton, see for example Ref. \cite{prim:string}. On
the other hand, in the mechanism we present above, we assume that
primordial magnetic fields may be created as the minimum of the one
loop effective energy. However the magnetic fields that obtain from
this mechanism, if we take into account the evolution of the early
universe and the conservation of the magnetic flux (for details see
Ref. \cite{EnqOle:1}), can not give the correct size of the seed
fields which are responsible for the generation of the present days
observed galactic magnetic fields.

Finally, we emphasize that a construction of a reliable scenario for
the generation of primordial magnetic fields is beyond the scope of
this paper. Our main motivation is to investigate a classical topic
like the Heisenberg Euler lagrangian in the case of extra
dimensions, and to note the rapid power growing of the quantum
energy in this case.

\section{Acknowledgements}
We would like to thank K. Tamvakis and A. Kehagias for reading and
comments on the manuscript. We also thank C. Bachas,  G.
Koutsoumbas, A. Lahanas, G. Savvidy and G. Tiktopoulos for useful
discussions. K. Farakos thanks CERN for hospitality during the last
stages of this work. The work of P.P. was supported by the
"Pythagoras" project of the Greek Ministry of Education.


\begin{thebibliography}{99}

\bibitem{1} W. Heisenberg and H. Euler, Z. Physik 38, 314 (1936).

\bibitem{2} V. Weisskoph, K. Dan. Vidensk. Selsk., Mat.-Fys
Medd. XIV, No. 6 (1936).

\bibitem{3} J. Schwinger, Phys. Rev. 82, 664 (1951).

\bibitem{4} A. N. Redlich, Phys. Rev. D 29, 2366 (1984).

\bibitem{6} G. Dunne and T. Hall, Phys. Lett. B 419, 322 (1998).

\bibitem{7} M. Bordag and K. Kirsten, Phys. Rev. D 60, 105019
(1999); M. Bordag and I. Drozdov, Phys. Rev. D68, 065026 (2003).

\bibitem{8} P. Pasipoularides, Phys. Rev.
 D 64, 105011 (2001); P. Pasipoularides, Phys. Rev. D 67, 107301 (2003).

\bibitem{9} H. Gies, K. Langfeld, Nucl.Phys., B613:353-365 (2001);  K. Langfeld, L.
Moyaerts,  H. Gies,  Nucl.Phys. B646:158-180 (2002).

\bibitem{11} H.W. Lee, P.Y. Pac and H.K. Shin,
Phys.Rev. D 40, 4202 (1989); V.P. Gusynin and I.A. Shovkovy, J.
Math. Phys. 40, 5406 (1999).

\bibitem{A11}I. Antoniadis, Phys. Lett. B 246 (1990) 377; I.
Antoniadis and K. Benakli, Phys. Lett. B326 (1994).

\bibitem{D11} N. Arkani-Hamed, S. Dimopoulos and G. R. Dvali, Phys.
Rev. D59, 086004 (1999); N. Arkani-Hamed, S. Dimopoulos and G. R.
Dvali, Phys. Lett. B429, 263 (1998).

\bibitem{12} K. R. Dienes, E. Dudas and T. Gherghetta, Nucl. Phys.
B537, 47 (1999); \textit{TeV-Scale GUTs}, K. R. Dienes, E. Dudas and
T. Gherghetta, hep-ph/9807522.

\bibitem{B12} C. P. Bachas, JHEP 11, 023 (1998).

\bibitem{Z12} J. Kubo, H. Terao, G. Zoupanos, Nucl. Phys. B574 (2000)
495-524.

\bibitem{13} C. P. Korthals Altes and M. Laine, Phys. Lett B511:269
(2001); C. P. Korthals Altes, \textit{Cosmological constraints from
extra-dimension induced domainwalls}, [hep-ph/0307368]; K. Farakos,
P. de Forcrand, C. P. Korthals Altes, M. Laine and M. Vettorazzo,
Nucl. Phys. B655:170 (2003); \textit{Effective potential analysis
for 5D SU(2) gauge theories at finite temperature and radius}, K.
Farakos and P. Pasipoularides, Nucl. Phys. B705 (2005) 92-110.

\bibitem{extra1} S. P. Gavrilov and D.M. Gitman, Phys. Rev. D53, 7162 (1996); E. V. Gorbar,
Phys. Lett. B491 (2000) 305-310.

\bibitem{14} T. Appelquist, Hsin-chia Cheng, B. A. Dobrescu, Phys.
Rev. D64, 035002 (2001).

\bibitem{peskin:14} \textit{An Introduction to Quantum field
Theory}, M. E. Peskin and D. V. Schroeder, Addison-Wesley Puplishing
Company (1995).

\bibitem{Houang:13} \textit{Quarks Leptons and Gauge Fields}, K. Huang, Word
Scientific (1992), 2nd Edition.

\bibitem{Wilson:1} J. F. Oliver, J. Papavassiliou, and A.
Santamaria, Phys. Rev. D67, 125004 (2003).

\bibitem{18} W. Dittrich, W. Tsai, K. H. Zimmermann, Phys. Rev. D19, 2929 (1979).

\bibitem{15} \textit{Quantum Electrodynamics}, W. Greiner and J.
Reinhardt, Springer-Verlag (1992).

\bibitem{Savvidy:1} S. G. Matinyan and G. K. Savvidy, Nucl. Phys.
B134, 539 (1978); G. K. Savvidy, Phys.Lett.B71, 133 (1977).

\bibitem{19} T. Vachaspati, Phys. Lett. B 265 ,258 (1991).

\bibitem{17}\textit{Electroweak Magnetism, W-condensation and Anti-Screening}, J. Ambj${\o}$rn and P. Olesen, hep-ph/9304220.

\bibitem{mag:sup} \textit{A way to break supersymmetry}, C. Bachas,
hep-th/9503030.

\bibitem{prim:1} \textit{Primordial Magnetic Fields and their
development}, P. Olesen, hep-ph/9708320.

\bibitem{prim:string} M. S. Turner and L. M. Widrow, Phys. Rev. D37, 2743 (1988);
M. Gasperini, M. Giovannini and G. Veneziano, Phys. Rev. Lett.
75:3796 (1995).

\bibitem{EnqOle:1} K. Enqvist and P. Olesen, Phys. Lett. B 329, (1994)
195-198.

\end{thebibliography}
\end{document}